\documentclass[aps,prb,superscriptaddress,preprintnumbers,
showpacs,letterpaper,twoside,twocolumn,amsmath,amssymb]{revtex4}
\usepackage{graphicx}     
\begin{document}

\title{Electronic and Magnetic Properties of $\rm{RMnO}_3/\rm{AMnO}_3$ Heterostructures}
\author{Rong Yu}\affiliation{Department of Physics and Astronomy,
University of Tennessee, Knoxville, TN 37996, USA} \affiliation{Materials
Science and Technology Division, Oak Ridge National Laboratory, Oak Ridge, TN 37831, USA}
\author{Seiji Yunoki} \affiliation{Computational Condensed Matter
Physics Laboratory, RIKEN, Wako, Saitama 351-0198, Japan}
\affiliation{CREST, Japan Science and Technology Agency (JST),
Kawaguchi, Saitama 332-0012, Japan}
\author{Shuai Dong}
\affiliation{Department of Physics and Astronomy, University of
Tennessee, Knoxville, TN 37996, USA} \affiliation{Materials Science and
Technology Division, Oak Ridge National Laboratory, Oak Ridge, TN 37831, USA}
\affiliation{Nanjing National Laboratory of Microstructures, Nanjing
University, Nanjing 210093, China}
\author{Elbio Dagotto}\affiliation{Department of Physics and Astronomy,
University of Tennessee, Knoxville, TN 37996, USA} \affiliation{Materials
Science and Technology Division, Oak Ridge National Laboratory, Oak Ridge, TN 37831, USA}
\date{\today}
\begin{abstract}
The ground state properties of $\rm{RMnO}_3/\rm{AMnO}_3$
($\rm{RMO/AMO}$) heterostructures (with R=La, Pr, ..., a trivalent
rare-earth cation, and A=Sr, Ca, ..., a divalent alkaline cation) are studied using a
two-orbital double-exchange model including the superexchange
coupling and Jahn-Teller lattice distortions. To describe
the charge transfer across the interface, the long-range Coulomb
interaction is taken into account at the mean-field level, by
self-consistently solving the Poisson's equation. The calculations are
carried out numerically on finite clusters. We find that the
state stabilized near the interface of the heterostructure is
similar to the state of the bulk compound $\rm{(R,A)MO}$ at
electronic density close to $0.5$. For instance, a charge and orbitally
ordered CE
state is found at the interface if the
corresponding bulk $\rm{(R,A)MO}$ material is a narrow-to-intermediate
bandwidth manganite. But instead the interface regime accommodates an A-type
antiferromagnetic state with a uniform $x^2-y^2$ orbital order, if
the bulk $\rm{(R,A)MO}$ corresponds to a wide bandwidth manganite.
We argue that
these results explain some of the properties of long-period
$\rm{(RMO)}_m\rm{/(AMO)}_n$ superlattices, such as
$\rm{(PrMnO_3)}_m/\rm{(CaMnO_3)}_n$ and
$\rm{(LaMnO_3)}_m/\rm{(SrMnO_3)}_n$.
We also remark that the intermediate states
in between the actual interface and the bulk-like
regimes of the heterostructure are dependent on the
bandwidth and the screening of the Coulomb interaction. In these
regions of the heterostructures, states
are found that do not have an analog in experimentally known bulk phase diagrams.
These new states of the heterostructures
provide a natural interpolation between magnetically-ordered states that
are stable in
the bulk at different electronic densities.
\end{abstract}

\maketitle

\section{Introduction}

Modern fabrication technology allows for the growth of multilayer
structures that are nearly perfect at the atomic level, namely with
minimal roughness, employing a variety of transition metal oxides
(TMO). Due to the simultaneous participation of several degrees of
freedom in TMO, it is expected that the artificial multilayer
structures made of these materials will exhibit much richer physics
than in conventional semiconductor heterostructures. Indeed, recent
studies have revealed some fascinating phenomena, such as the
reconstruction of spin, charge, and orbital orders at the
interface.~\cite{Chakhalian08,Izumi99} Many interesting properties
have also been theoretically predicted using a variety of many-body
techniques.~\cite{Okamoto04,Chaloupka08,Yunoki07}

Among  the several ongoing efforts, there is a considerable interest in the
analysis of
$\rm{RMnO}_3/\rm{AMnO}_3$ ($\rm{RMO/AMO}$) heterostructures, where
R=La, Pr, ... refers to a trivalent rare-earth, and
A=Sr, Ca, ... is a divalent alkaline
element.~\cite{Lin08,Adamoetal08,Yamadaetal06,Nanda08,May07,
Smadici07,Koida02,Bhattacharya07,DongLMOSMO08} At low temperatures,
the bulk $\rm{RMO}$ is in an
A-type antiferromagnetic (A-AFM) state, which is an
insulator, whereas the bulk $\rm{AMO}$ is in a G-type antiferromagnetic
(G-AFM) state, that is also an insulator. Upon doping,
the alloy $\rm{R_{1-{\it x}}A_{\it x}MnO_3}$
($\rm{(R,A)MO}$) exhibits a variety of states depending on the doping
concentration $x$, which controls the charge density in the alloy.
However, the heterostructure $\rm{RMO/AMO}$ could potentially behave differently from
its parent bulk compounds. For instance, the transfer of
charge through the
interface caused by the different Fermi energies, and concomitant different electronic density
concentrations, of the superlattice components causes
a distribution of charge that it is not homogeneous along the
growth direction. Hence, several states may exist in different
regions of the heterostructure. While far from the
interface the behavior must be similar to the one in the bulk
compounds, close to the interface it may occur that phases very different
from those of the superlattice components
may exist due to the charge leaking through the interfaces. For
instance, in the short-period $\rm{LaMnO_3/SrMnO_3}$
($\rm{LMO/SMO}$) superlattices the regime close to the interface
exhibits ferromagnetic (FM) metallic behavior, which is different
from either $\rm{LaMnO_3}$ or
$\rm{SrMnO_3}$.~\cite{Smadici07,Adamoetal08,May07,Bhattacharya08}
More interestingly, as the number of $\rm{Sr}$ layers exceeds a
critical value, the metallic behavior gives way to an insulating
one, displaying a metal-to-insulator transition (MIT). This suggests
that the electronic reconstruction at the interface has a crucial
effect on the physical properties of the heterostructure.\cite{Okamoto04}
Theoretically, considerable progress has been made in describing the
ferromagnetism induced by this electronic
reconstruction.~\cite{Nanda08,Lin06,Yunoki08,Brey07,Brey08,Lin08} A recent
study by the authors~\cite{DongLMOSMO08} focused on the MIT in the
short-period $\rm{LMO/SMO}$ superlattices. There, a FM metallic state
with a dominant $3z^2-r^2$ orbital order was found at the interface. The
insulating behavior of the superlattice was explained as induced by
Anderson localization of this quasi two-dimensional (2D) FM state.
Note that limited by computing power, most theoretical
efforts based on computer simulations
concentrate on superlattices with a small number of
$\rm{Sr}$ layers. This is enough to understand the FM in the
short-period superlattice, but may not be appropriate to study the
properties of longer-period structures. Another limitation is that
most theoretical studies focus on the $\rm{LMO/SMO}$ superlattice. The
corresponding bulk compound $\rm{(La,Sr)MnO_3}$ ($\rm{LSMO}$) is a
wide bandwidth manganite. But not much is known about the interfacial
state of the
$\rm{RMO/AMO}$ superlattice if $\rm{(R,A)MO}$ is a narrow or
intermediate bandwidth manganite.

In this manuscript, we use the two-orbital double-exchange model for
manganites, which has been successfully applied to the study of bulk
Mn oxides,~\cite{DagottoReview01} to the analysis of $\rm{RMO/AMO}$
heterostructures. These heterostructures are assumed to be grown along
the $(0,0,1)$ direction ($c$ axis).
We here also assume that the length of the heterostructure is long enough
that the states at the two ends of the heterostructure resemble
their bulk counterparts, i.e., an A-AFM at the RMO side and a G-AFM
at the AMO side. This assumption allows us to focus on the only
interface present in the heterostructure. By using a relaxation method
introduced in the following section, we can obtain the magnetic and
electronic properties of the ground state of the heterostructure.
Different from previous efforts that concentrated on the FM
tendency in short-period superlattices, we find that in the
heterostructure considered in this paper the state stabilized at the
interface depends on the bandwidth of the corresponding bulk
compound $\rm{(R,A)MO}$ at electronic density close to $0.5$. If
$\rm{(R,A)MO}$ is a narrow-to-intermediate bandwidth manganite,
a state resembling the well-known CE-state of bulk manganites
 is found to be stabilized at the interface. Hence our calculations suggest a
CE-like interface state in the $\rm{PrMnO_3}/\rm{CaMnO_3}$
heterostructure.~\cite{commentbrey} But if $\rm{(R,A)MO}$ is a wide bandwidth
manganite, an A-AFM state is found at the interface instead. This is
consistent with recent experimental results on the
$\rm{(LMO)}_n/\rm{(SMO)}_{2n}$ superlattices.~\cite{Bhattacharya07}

In addition, it is important to remark that one of the main results of
our study is the observation of states close to the interface
that do $not$ have an analog in experimentally known bulk phase diagrams.
These states arise as interpolations
between, e.g., the A-AFM and CE-states that dominate in the bulk and
interfaces, respectively.
For instance, canting of the spins in the CE zigzag chains creates
a novel ``canted CE state''. Also the relative spin angle between adjacent
CE planes can be different from those observed in the bulk. And in some occasions,
arrangements of spins and orbitals were identified that do not have a
clear bulk analog in experiments.


This paper is organized as follows. In Sec.~II, we
introduce the two-orbital model and the numerical method used to
obtain the ground state properties of the heterostructure. The
results for $\rm{(R,A)MO}$ corresponding to narrow-to-intermediate
bandwidth manganites are presented in Sec.~III. Results for
$\rm{(R,A)MO}$ corresponding to wide-bandwidth manganites are
provided in Sec.~IV. Finally, in Sec.~V a discussion of our
results is given, followed by conclusions.

\section{Model and Numerical Method}

To investigate the physical properties of the
$\rm{RMO/AMO}$ heterostructure described in Sec.~I, here
the two-orbital model for
manganites~\cite{DagottoReview01} will be applied. This model has been
widely used before to study the properties of bulk manganites, and
the following assumptions are also widely accepted: (1) the $t_{\rm 2g}$ electrons are
considered as localized and are described as classical spins with magnitude
$S=3/2$; (2) the
Jahn-Teller lattice distortions are also assumed to be classical; (3) the Hund
coupling between the $t_{\rm 2g}$ and $e_{\rm g}$ electrons is assumed to be
infinitely large so that the $e_{\rm g}$ spin is always
parallel to the localized $t_{\rm 2g}$ spin at the same site.
Based on these assumptions, the Hamiltonian that will be  applied to the
manganite heterostructure reads
\begin{eqnarray}\label{E.Ham}
\nonumber H &=& -\sum_{<i,j>}^{\alpha,\beta} \left(
t_{\textbf{r}}^{\alpha\beta} \Omega_{ij} c_{i\alpha}^{\dagger}
c_{j\beta}+\rm{H.c.}\right) + \sum_{<i,j>}
J^{\rm{AF}}_{i,j} \textbf{S}_{i} \cdot \textbf{S}_{j} \\
\nonumber &+& \sum_{i}(\phi_i+W_i) n_{i}+\lambda \sum_{i}
(Q_{1i} n_{i}+Q_{2i} \tau^x_i + Q_{3i} \tau^z_i) \\
&+& \frac{1}{2} \sum_{i} (2Q_{1i}^2 + Q_{2i}^2 + Q_{3i}^2).
\end{eqnarray}
The first term of Hamiltonian Eq.(1) denotes the two-orbital
double-exchange hopping term. $\alpha$ and $\beta$ run over the two $e_{\rm g}$
orbitals $d_{\rm x^2-y^2}$ (orbital $a$) and $d_{\rm 3z^2-r^2}$ (orbital
$b$) of a Mn ion. $c_{i\alpha}$ ($c_{i\alpha}^{\dagger}$)
annihilates (creates) an $e_{\rm g}$ electron in orbital $\alpha$ at
site $i$ with its spin parallel to the localized $t_{\rm 2g}$ spin at site
$\textbf{S}_{i}$. \textbf{r} denotes the exchange direction, giving
$t_{x}^{aa}=t_{y}^{aa}=3t_{x}^{bb}=3t_{y}^{bb}=3t_0/4$,
$t_{y}^{ab}=t_{y}^{ba}=-t_{x}^{ab}=-t_{x}^{ba}=\sqrt{3}t_0/4$,
$t_{z}^{aa}=t_{z}^{ab}=t_{z}^{ba}=0$ and $t_{z}^{bb}=t_0$ ($t_0$ is
set to be the energy unit). The hopping amplitude is affected by the factor $\Omega_{ij} =
\cos(\frac{\theta_{i}}{2}) \cos(\frac{\theta_{j}}{2}) +
\sin(\frac{\theta_{i}}{2}) \sin(\frac{\theta_{j}}{2})
\exp[-i(\varphi_{i}-\varphi_{j})]$, where $\theta_i$ and $\varphi_i$
are the angles of the $t_{\rm 2g}$ spins in spherical coordinates. Here, we
will assume that the hopping amplitudes are the same for electrons on both
sides of the heterostructure. This may not be realistic given the
possible mismatch of lattice constants between the two different
compounds. However, since the superexchange coupling is more
sensitive to the change of lattice constant than the hopping amplitudes
themselves, as a first
approximation we assume that the hopping constants keep the same value on
both sides while the superexchange couplings may become layer
dependent. The second term is the standard superexchange interaction between
nearest-neighbor (NN) $t_{\rm 2g}$ spins. Here the $t_{\rm 2g}$ spin
$\textbf{S}_{i} = (\sin\theta_i\cos\varphi_i,
\sin\theta_i\sin\varphi_i, \cos\theta_i)$ has been normalized to a
unit vector (the actual $S$=3/2 magnitude of the spins is absorbed in
the superexchange coupling). To consider the effect of possible distortions from
a perfect cubic lattice, two couplings are used:
$J^{\rm{AF}}_{i,j}=J^{\rm{AF}}_\parallel$
if $i$ and $j$ are NN sites in the same layer (with same $z$
coordinate), and $J^{\rm{AF}}_{i,j}=J^{\rm{AF}}_\perp$ if $i$ and
$j$ are NN sites belonging to two adjacent layers. In the third
term, $\phi_{i}$ corresponds to a site-dependent Coulomb potential
that originates from the charge transfer through the interface, and it is
determined via the Poisson equation as described below.
$W_i$ denotes the work function on either side, which is determined
by the positions of the chemical potentials in the corresponding bulk
materials. More details on $\phi_{i}$ and $W_i$ will be discussed
later in this section. $n$ is the $e_{\rm g}$ charge density. The fourth
term stands for the electron-phonon coupling. The
$Q$s are lattice distortions for
the Jahn-Teller modes ($Q_{2}$ and $Q_{3}$) and breathing mode ($Q_{1}$).
$\mathbf{\tau}=(\tau^x,\tau^y,\tau^z)$ is the orbital pseudospin
operator. The last term is the elastic energy of the lattice distortions
considered here. The extra factor 2 for the breathing mode suppresses this
mode as compared with the Jahn-Teller modes that are the most active.\cite{Hotta99}

\begin{figure}
\begin{center}
\vskip -0.5cm
\includegraphics[
bbllx=0pt,bblly=0pt,bburx=546pt,bbury=420pt,
width=75mm,angle=0]{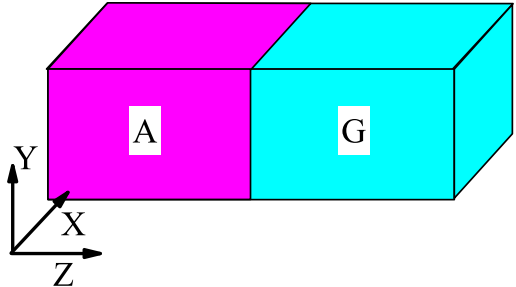} \caption{Schematic representation of
the heterostructure studied in this paper. It is defined on a
$4\times4\times8$ lattice. The bulk material on one side of the
heterostructure is in the A-AFM phase, while the bulk material on
the other side is in the G-AFM phase.} \label{F.Heterostruct}
\end{center}
\end{figure}

For any given $t_{\rm 2g}$-spin and lattice configuration the above
Hamiltonian can be solved by numerically diagonalizing the bilinear fermionic
sector. The ground state is approached via a relaxation technique:
the optimized configuration of the oxygen coordinates
and the $t_{\rm 2g}$ spins is
determined by minimizing the total energy of the Hamiltonian
Eq.(\ref{E.Ham}). This method is first applied to a
$4\times4\times4$ lattice with periodic boundary conditions to
estimate properties of the bulk compound on each side of the
heterostructure. In this publication, we will focus on the interface
between an A-AFM and a G-AFM. Hence, we will adopt a set of parameters
$J^{\rm{AF}}$ and $\lambda$ that gives A-AFM and G-AFM ground states
at the limits of $e_{\rm g}$ electron densities $n=1$ and $n=0$, respectively. The same
set of parameters is then used to calculate the ground state of the
heterostructure. The heterostructure is defined on a
$4\times4\times8$ lattice (see Fig.~\ref{F.Heterostruct}), with
periodic boundary conditions in the $xy$ plane ($ab$ plane) and open
boundary conditions along the $z$ direction ($c$ axis). The initial
spin configuration is set to be A-AFM on one side of the
heterostructure ($4$ layers) with electron density $n=1$, and G-AFM
on the other side ($4$ layers) with $n=0$. Regarding the relative spin
direction between the A-AFM and G-AFM states, their relative angles
$\theta$ and $\phi$ were allowed to take 16 equally-spaced values
in their respective ranges, i.e. [0,$\pi$] for $\theta$ and [0,2$\pi$] for $\phi$,
thus giving 256 possibilities. For each
of these 256 possibilities,
and a fixed set of couplings in the Hamiltonian,
an independent optimization of the classical variables
was made, namely
the energy was minimized by solving
\begin{equation}\label{E.Minimize}
\nabla_{\psi_i}\langle H_{2b}\rangle = 0
\end{equation}
self-consistently using the Broyden's method,~\cite{NumRec} where
$\psi_i=(\theta_i,\varphi_i,Q_{1i},Q_{2i},Q_{3i})$. At each step we
keep the two end layers to be dominated by the A-AFM and G-AFM states,
respectively. Namely, the dominant layer wavevector for the
Fourier transform of the spin-spin correlations must be (0,0) and
($\pi$,$\pi$), respectively, otherwise the configuration is discarded.
The rest of the layers, of course, may have different magnetic orders. The
optimized spins and oxygen coordinates of the ground state then
corresponds to the configuration with the lowest overall energy
after this long optimization process.

The charge transfer through the interface is taken into account via
the self-consistent solution of the Coulomb potential $\phi_i$.\cite{Okamoto04}
For the heterostructure discussed here, involving
manganites only, we assume $W_{\rm A}=W_{\rm G}$ and make them equal to
zero for simplicity. This approximation is in agreement with results of
previous discussions.\cite{Yunoki07}
Then, the charge
transfer through the interface is fully driven by the charge density
difference, i.e., the charge always transfers from the A-AFM side to the G-AFM
side. To properly describe the charge transfer, the long-range Coulomb
interaction among the mobile $e_{\rm g}$ electrons and the positively charged
ionic background must be included into the Hamiltonian as
\begin{eqnarray}\label{E.HCoul}
H_{\rm{Coul}} &=& \alpha t_0\sum_{i\neq j} \left[\frac{1}{2}
\frac{n_i n_j}{|\vec{r}_i-\vec{r}_j|} - \frac{n_i
n^+_j}{|\vec{r}_i-\vec{r}_j|}\right],
\end{eqnarray}
where $\alpha=e^2/\epsilon at_0$ is a dimensionless screening
parameter. For manganites, it is known that
$t_0$ is of the order of 0.5~eV and the lattice constant
$a\approx4\text{\AA}$, hence $\alpha$ depends on the choice for the dielectric
constant $\epsilon$. However, $\epsilon$ is both temperature
and frequency dependent and for this reason an accurate estimation
of $\alpha$ is
not well known. In this paper, the Hamiltonian
Eq.(\ref{E.HCoul}) is studied over a broad range of $\epsilon$ values,
$2\leqslant\epsilon\leqslant20$, corresponding to
$0.2\leqslant\alpha\leqslant2$. $\vec{r}_i$ is the position vector
of the  Mn site $i$. $n_i$ is the local electronic density at site
$i$. $n^+_i$ stands for the effective positive charge density on the
$i$-th Mn site arising from the background ions. Note that to simplify the
model, we have already assumed that all the charges from the background
ions are located on the Mn sites. Therefore, for the $\rm{RMO/AMO}$
heterostructure, $n^+_i$ is fixed to $1$ at the A-AFM side, and to
$0$ at the other side to enforce the charge neutrality.

Equation~\ref{E.HCoul} is solved at the mean-field level by introducing
the Coulomb potential
\begin{equation}
\phi_i = \sum_{j\neq i} \frac{\langle
n_j\rangle-n^+_j}{|\vec{r}_i-\vec{r}_j|}.
\end{equation}
We then find that the Coulomb interaction in Eq.(\ref{E.HCoul})
recovers the third term in Eq.(\ref{E.Ham}). In practice, the Coulomb
potential $\phi_i$ is determined at each step of the relaxation
procedure by self-consistently solving the Poisson's equation
\begin{equation}\label{E.Poisson}
\nabla^2 \phi_i = \alpha \left(\langle n_i\rangle - n^+_i\right).
\end{equation}
Numerically the following discretization is applied:
$\partial^2\phi/\partial x^2 = \phi_{i+\hat{x}} - 2\phi_{i} +
\phi_{i-\hat{x}}$, $\partial^2\phi/\partial y^2 = \phi_{i+\hat{y}} -
2\phi_{i} + \phi_{i-\hat{y}}$, and $\partial^2\phi/\partial z^2 =
\phi_{i+2\hat{z}} - 2\phi_{i+\hat{z}} + \phi_{i}$. The open boundary
condition is applied along the $z$ direction (out-of-plane) but
periodic boundary conditions are applied along the $x$ and $y$ directions
(in-plane).

In this work, up to 4,000 iterations per set of couplings (an
iteration here is defined as an update of the entire set of
classical variables in the cluster) are used to obtain the optimized
spin and oxygen lattice configurations of the ground state. The most
typical number of iterations is approximately 1,000. At each
iteration of this relaxation procedure, up to 2,000 additional
iterations at a fixed set of classical variables are used to solve
the Poisson's equation. Note that the diagonalization of fermions
must be performed at each step in solving this Poisson's equation.
Then, in order to find the optimized configuration for the ground
state on a $4\times4\times8$ lattice, typically approximately $10^6$
times the exact diagonalization of the $256\times256$ matrix is
necessary. This is very CPU time demanding. It is for these
practical reasons that only the $4\times4\times8$ lattice is used
here to study the properties of the heterostructure.

\section{Results for narrow to intermediate bandwidth
manganites}\label{S.LMM}

In this section, results for the RMO/AMO heterostructure will be discussed,
where (R,A)MO corresponds to a narrow-to-intermediate bandwidth
manganite, such as $\rm{(La,Ca)MnO_3}$ and $\rm{(Pr,Ca)MnO_3}$.
Therefore, the results presented here are expected to best describe the properties of
$\rm{LaMnO_3/CaMnO_3}$ or $\rm{PrMnO_3/CaMnO_3}$ heterostructures.

\subsection{Phase diagram of the bulk material}
Before discussing the properties of the heterostructure, we will
first analyze the magnetic phase diagram of the corresponding bulk
material. On one hand, this allows us to determine the model
parameters to be used for the calculation of heterostructures; on the other
hand, a better knowledge of the bulk phase diagram also helps in
understanding the possible spin structures in the heterostructure.

\begin{figure}
\begin{center}
\includegraphics[
clip,width=90mm,angle=0]{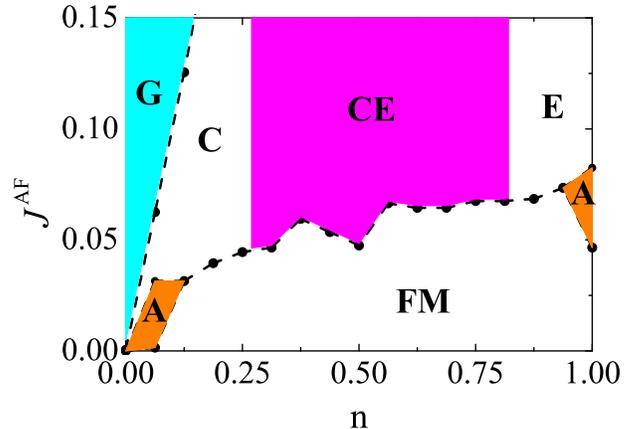} \caption{Phase diagram of the
two-orbital model on the $4\times4\times4$ cubic lattice with
$J^{\rm{AF}}_\parallel=J^{\rm{AF}}_\perp=J^{\rm{AF}}$ and
$\lambda=1.5$. The results are obtained comparing the energies of
the G-AFM, C-AFM, CE-AFM, E-AFM, A-AFM, and FM states.}
\label{F.PhDCE}
\end{center}
\end{figure}

Previous theoretical investigations~\cite{DagottoReview01} have shown that
in the two-orbital model the bandwidth depends on the
electron-phonon coupling strength $\lambda$ and the superexchange
coupling strength $J^{\rm{AF}}$. The larger the $\lambda$ and
$J^{\rm{AF}}$ are, the narrower the bandwidth is. In
Fig.~\ref{F.PhDCE} the phase diagram of the two-orbital
model for the bulk is shown at various electronic densities, using $\lambda=1.5$
and $J^{\rm{AF}}_\parallel=J^{\rm{AF}}_\perp=J^{\rm{AF}}$. The phase
diagram is obtained by comparing energies of several candidate
states: A-AFM, G-AFM, C-AFM, E-AFM, CE-AFM, and FM states on the
$4\times4\times4$ lattice. Note that there could exist even more exotic
states in the phase diagram,
such as the $\rm {C_{\it x}E_{1-{\it x}}}$ state previously proposed\cite{Hotta03}
at $n>0.5$ if larger lattice sizes could be considered,
and spiral states at $n=1$ when in the presence of spin frustration.\cite{spiral}
However, as shown below our interest will be  mainly in the A-AFM state
stabilized at electronic density $n=1$, the CE and A-AFM states at $n=0.5$, and the G-AFM state
at $n=0$, namely in regions where the
$\rm{C_{\it x}E_{1-{\it x}}}$ and spiral states are not expected to be relevant.
Hence in the
current study, the $\rm{C_{\it x}E_{1-{\it x}}}$ and spirals states are not considered.

The bulk phase diagram is very rich. At small superexchange
couplings the ground state is FM in a broad density regime, as expected
from the double-exchange mechanism. At
larger superexchange coupling the system transitions, from low to high electronic
densities, from G-AFM, to C-AFM, to CE-AFM, and finally to E-AFM phases, respectively. Close to
$n=0.5$ the CE phase is stable over a wide range of $J^{\rm{AF}}$
values and has an alternate charge/orbital order; whereas the G-AFM
phase at low $n$ has neither charge nor orbital order. There are
also two A-AFM phases: close to electronic density $n=1$ at
intermediate $J^{\rm{AF}}$ and close to $n=0$ at small $J^{\rm{AF}}$ values.
The one close to $n=1$ has the correct alternate $3x^2-r^2$/$3y^2-r^2$
orbital order expected at $n=1$ from experimental information.~\cite{Hotta99,bulkAphase}
Thus, it is interesting to point out that in the range $0.05\leqslant
J^{\rm{AF}}\leqslant 0.075$ the phase diagram consists of G-AFM,
C-AFM, CE, FM, and A-AFM phases consecutively from low to high
electronic densities, correctly resembling the phase diagram of real
narrow-to-intermediate bandwidth bulk
manganites.~\cite{Kajimotoetal02} Hence,
$J^{\rm{AF}}_\parallel=J^{\rm{AF}}_\perp=0.065$ and $\lambda=1.5$
will be used as couplings
for calculations of the heterostructures to be discussed later in this section.

\subsection{CE state close to the interface of the heterostructure}
\begin{figure}
\begin{center}
\includegraphics[
clip,width=90mm,angle=0]{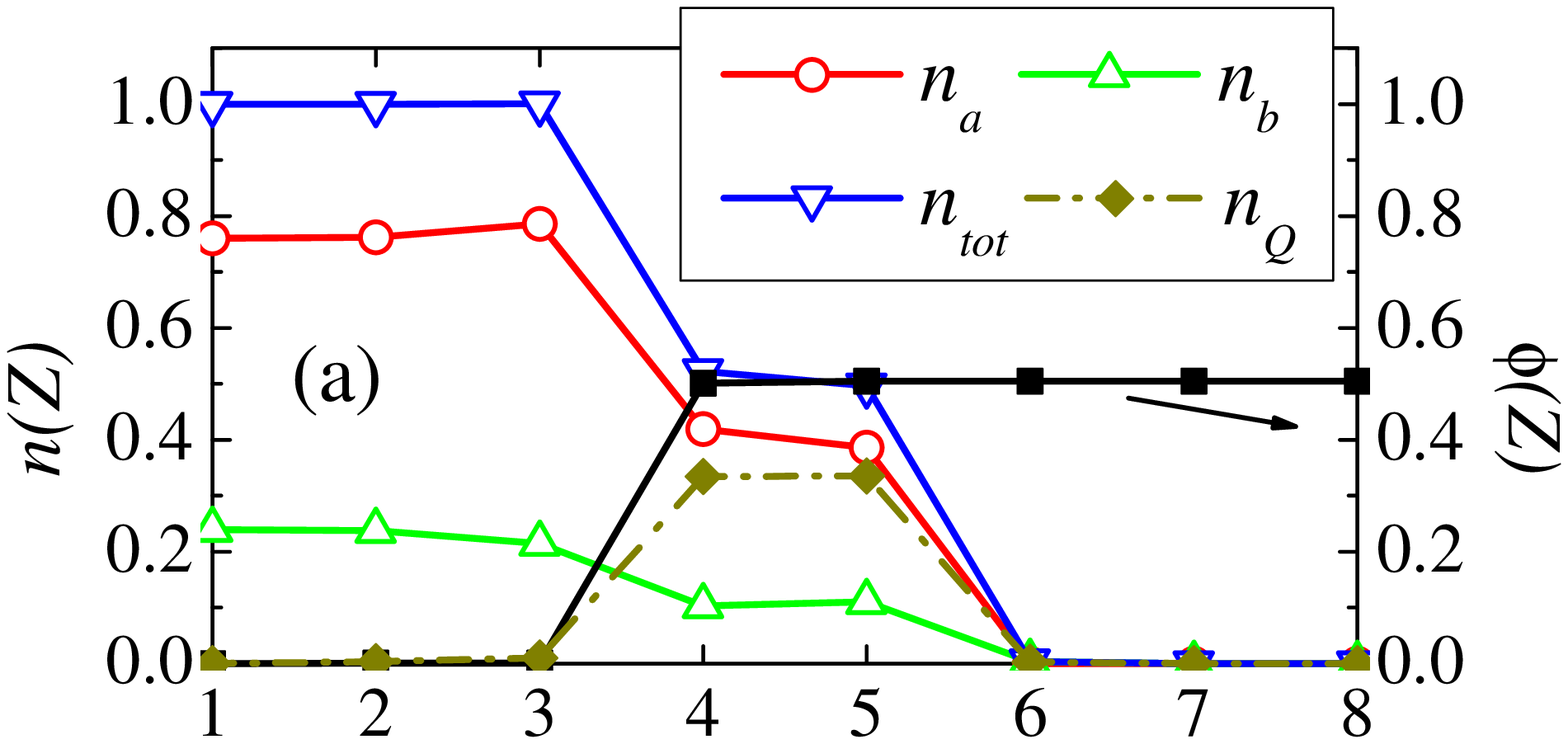} \vskip -1.cm
\includegraphics[
clip,width=90mm,angle=0]{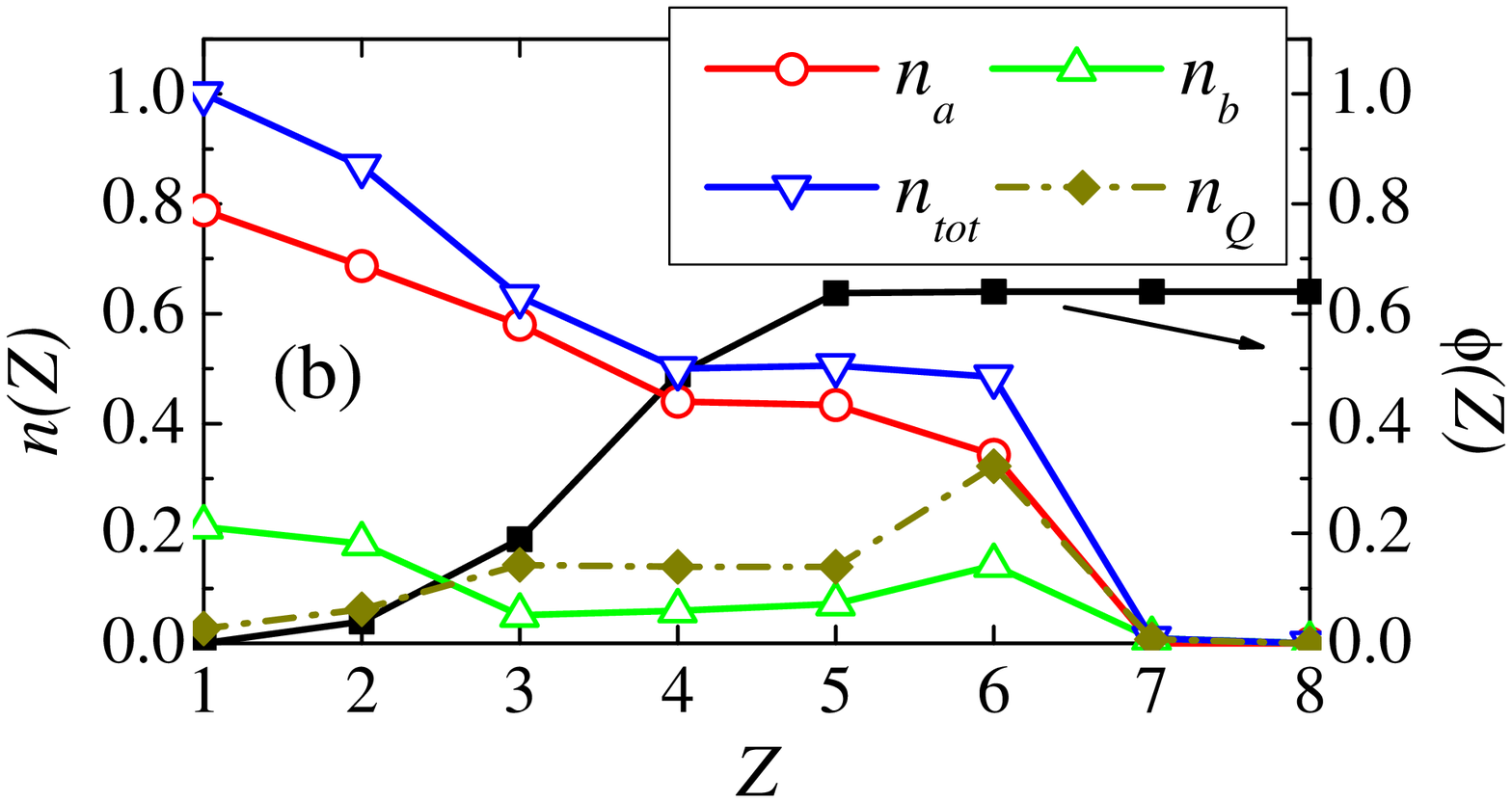} \caption{Layer-averaged
electronic density $ n(Z)$ and electrostatic potential $\phi(Z)$, at
$\lambda=1.5$ and $J^{\rm AF}=0.065$, vs. layer index. $n_a$ and
$n_b$ refer to the electronic densities of the $a$ and $b$ orbitals,
and $n_{\rm tot}=n_a+n_b$, $n_\mathbf{Q}$ is the Fourier transform
of the local electronic density in each layer at
$\mathbf{Q}=(\pi,\pi)$. (a) are results at $\alpha=1.0$; (b) are
results at $\alpha=0.3$.} \label{F.LcdCE}
\end{center}
\end{figure}

\subsubsection{Emergence of CE properties near the interface}

In this subsection, the physical properties of the
heterostructure are investigated using the above described
model parameters. In Fig.~\ref{F.LcdCE},
the averaged electronic density $n(Z)$ and electrostatic
potential $\phi(Z)$ in each layer are presented at $\alpha=1.0$ and $\alpha=0.3$.
At the two end layers of the heterostructure, the densities $n(Z=1)\approx1$ and
$n(Z=8)\approx0$ converge to the expected values in the corresponding
bulk materials. But charges are redistributed in the rest of the layers due
to the long-range Coulomb interactions. Note that there exists a
plateau at $n_{\rm tot}(Z)\approx0.5$ in the layers close to the interface.
To understand this feature, let us study the Fourier transform of
the local electronic density in each layer,
$n_{\mathbf{Q}}=\frac{1}{N_{\rm XY}}\sum_i n_i
e^{i\mathbf{Q}\cdot\mathbf{r}_i}$, where $N_{\rm XY}$ is the number of
sites in each layer. $n_{\mathbf{Q}}$ at $\mathbf{Q}=(\pi,\pi)$
displays a peak in the layers close to the interface, where the plateau in
$n_{\rm tot}$ exists. This suggests the presence of a
charge ordered phase with
$n_{\rm tot}(Z)\approx0.5$ that is stabilized at the interface. Such a charge
ordered state is found to appear at the interface for all the $\alpha$
values  considered in our study. Hence, we believe the existence of this state
is an intrinsic property at the interface of the heterostructure, at least
within the approximations used in our calculations.

\begin{figure}
\begin{center}
\includegraphics[
clip,width=85mm,angle=0]{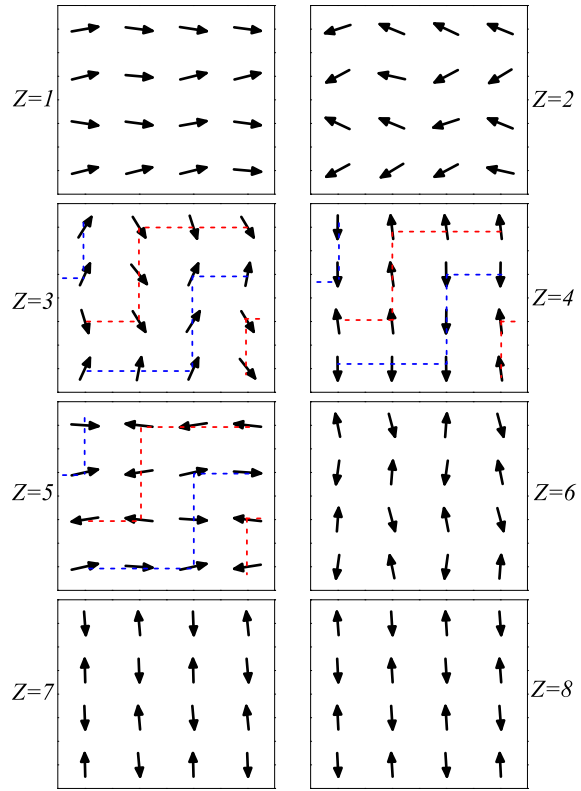} \caption{Optimized spin
configurations of the heterostructure studied here, using
$\lambda=1.5$, $J^{\rm{AF}}=0.065$, and $\alpha=1.0$. The dashed
lines highlight the spin zigzag chains that are characteristic of CE
states,  with FM order within each chain.} \label{F.SpnCE}
\end{center}
\end{figure}

To better understand the states that are located at or close to the
interface, it is important to study how the spins are ordered in
each layer of the heterostructure. We have observed that the spin
order is not much sensitive to the value of
$\alpha$.~\cite{note_spin_order} Thus, here only the information for
the optimized real-space spin patterns at $\alpha=1.0$ is presented
in Fig.~\ref{F.SpnCE}. From this figure, it is clear that the layers
close to the interface (layers 3, 4, and 5) exhibit a CE-type spin
order. It is well known that in bulk systems and narrow bandwidth
manganites, the CE state is stabilized at electronic density
$n\approx0.5$ and has a staggered charge order. Moreover, our
previous results in this section using a cubic cluster to mimic the
bulk also unveiled a CE state at the same density. Thus, the CE spin
order found here further confirms that a charge/spin/orbital ordered
state is stabilized at the interface.~\cite{commentbrey}

\begin{figure}
\begin{center}
\includegraphics[bbllx=0pt,bblly=0pt,bburx=585pt,bbury=455pt,
clip,width=85mm,angle=0]{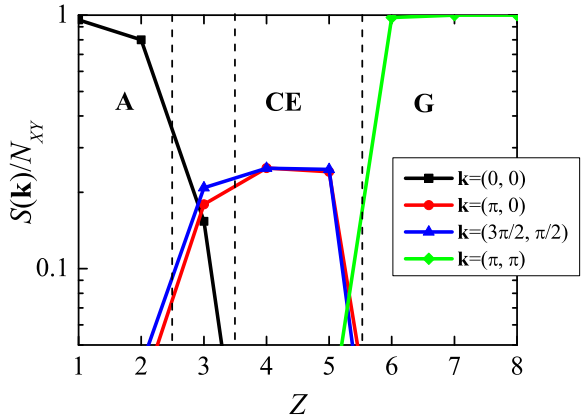} \caption{Layer dependence of the
spin structure factor $S(\mathbf{k})$ for the optimized spin
configurations shown in Fig.~\ref{F.SpnCE} at several momenta
$\mathbf{k}$: $(0,0)$, $(\pi,0)$, $(3\pi/2,\pi/2)$, and
$(\pi,\pi)$.} \label{F.StructCE}
\end{center}
\end{figure}

Moving from the interface
towards one end of the heterostructure (layer 1), the
CE state gives way to the FM order in each layer. But spins in two adjacent layers
tend to be AFM coupled, see layers 1 and 2 for instance.
Hence, the state at this end of the
heterostructure has an A-AFM tendency, resembling the spin order
in the bulk. Similarly, the state on the other side of the
heterostructure turns from the CE state to a G-AFM. The above
described
features are further confirmed by the layer-dependent spin structure
factor $S(\mathbf{k})=\frac{1}{N_{\rm XY}} \sum_{i,j}
\mathbf{S}_i\cdot\mathbf{S}_j
e^{i\mathbf{k}\cdot(\mathbf{r}_i-\mathbf{r}_j)}$ shown in
Fig.~\ref{F.StructCE}. Considering the degeneracy of states with
ordering vector $\mathbf{k}=(\pi,0)$ and $(0,\pi)$, and with
$\mathbf{k}=(\pi/2,3\pi/2)$ and $(3\pi/2,\pi/2)$, the CE order in
layers 4 and 5 is nearly perfect.

\subsubsection{Novel states at the interface
with no bulk analog in experimentally known phase diagrams}

The previous analysis shows that the interface has CE characteristics.
This may be considered as an ``obvious'' result, since in a heterostructure
the interpolation between bulk materials with $n=1$ and $n=0$ likely will induce
$n=0.5$ at the interface. In this simplistic conceptual framework, the properties at the interface
can be guessed, with good accuracy, merely from the bulk phase diagram. While
this provides a reasonable starting point to analyze results and make predictions,
further analysis actually suggests that this is not the end of the story, and
some surprises can be unveiled at interfaces.

To illustrate this point, consider for instance layer 3. Here, the average density
is very close to 1, yet the spins form zigzag chains as in the CE state of $n=0.5$.
The reason is that the spins at layer 3 are already being influenced by the robust
CE state formed at layer 4. Thus, layer 3 is an exotic interpolation between the
extremes case of the FM layers of the A-AFM state in one end, and the CE state at the
interface. Moreover, note that the relative orientation between the spins of adjacent zigzag
chains of layer 3 is not antiferromagnetic, as in a normal CE state, but it has some
canting. This ``canted CE state''
is not present in bulk phase
diagram, to our knowledge. Note that a state called the ``pseudo CE" was previously
discussed in experiments.\cite{cox} Considering just the individual layers, this
state is the same as the usual CE, but the coupling between CE layers is ferromagnetic, instead of antiferromagnetic
as in the standard CE state. Thus, this
pseudo CE state is not the same observed in our heterostructures.
Also note that a  ``canted CE'' state similar to that described in the present investigations was reported in early
neutron diffraction experiments\cite{yoshizawa} of $\rm{Pr_{0.7}Ca_{0.3}MnO_3}$. However, further
investigations for the same material\cite{cox,radaelli} suggested instead a mixed-phase interpretation of the
results, with a mixture of AF and FM phases, as opposed to a uniform canted CE state. Thus,
to our knowledge there is no evidence that the canted CE state exists in bulk form in real experiments, although
more work is needed to fully address this matter.
Also we are not aware of
previous theoretical investigations reporting such a canted CE state.~\cite{khomskii}

We also noticed
that although the layers located right at the interface, i.e. layers 4 and 5, exhibit almost a
perfect CE order individually, the spins in the two layers are not perfectly AFM
aligned with respect to one another compared with
what they should be in a bulk CE phase. In other words, the CE state
in the bulk is well known for showing a ``stacking'' property along the
direction perpendicular to the CE plane, which is respected here with
regards to charge and orbital but not with respect to the AFM relative order of the spins.
In fact, in the layers $Z$=4 and 5
of the heterostructure, the relative spin orientation is at approximately
90$^{\rm o}$ degrees, namely they are spin perpendicular to one another. Once again, to our
knowledge
such an arrangement does not exist in the bulk.
While it is obvious that the interplane spin deviation from AFM order in a CE arrangement
will increase
the superexchange energy of the system, these spin arrangements also allow for
charge transferring out-of-plane.  In other words, having an AFM order in a link reduces
the effective hopping amplitude to zero in a double-exchange context.
But if the spin order is not AFM, the kinetic
energy in that link improves.
Then, the system can gain double exchange energy in the $z$ direction
at the expense of superexchange
if the CE stacking does not involve AFM order. As a combined
effect, the total energy appears to be lowered by this mechanism
at the interfaces we studied here, while
in the bulk it does not occur.


\begin{figure}
\begin{center}
\vskip -0.5cm
\includegraphics[
clip,width=90mm,angle=0]{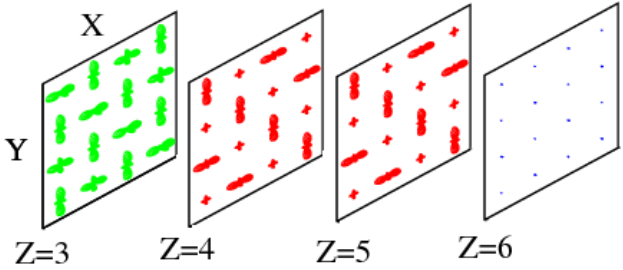} \vskip -0.8cm
\caption{Real-space orbital pattern of the $e_{\rm g}$ electrons for
the layers close to the interface. The model parameters are the same
as in Fig.~\ref{F.SpnCE}. The radian part of the electron wave
function (shown) is proportional to the local $e_{\rm g}$ electronic
density $n_i$. The plot displays a transition from the staggered
$3x^2-r^2$/$3y^2-r^2$ order in layer 3, to a CE-type orbital order
in layers 4 and 5, and then to a weak $3z^2-r^2$ order in layer 6,
consistent with the spin patterns in Fig.~\ref{F.SpnCE}.}
\label{F.OrbCE}
\end{center}
\end{figure}

The charge distribution in Fig.~\ref{F.LcdCE} suggests that on
average the $a$ ($x^2-y^2$) orbital has a higher occupation number
than the $b$ ($3z^2-r^2$) orbital. To obtain the exact orbital
pattern at each site, the expectation values of the local
pseudo-spin operators $\langle \tau^x_i \rangle$ and $\langle
\tau^z_i \rangle$ are calculated. Defining an effective phase angle
$\xi_i=\pi+\tan^{-1}(\langle \tau^x_i \rangle/\langle \tau^z_i
\rangle)$, we introduce a dressed state $|b\rangle =
\left[-\sin(\xi_i/2)c^\dagger_{ia} +
\cos(\xi_i/2)c^\dagger_{ib}\right] |0\rangle$ from which the
orbital occupation is computed as $\langle
b|n_i|b\rangle$. The details of this type of calculations
are well-known and they
can be found in Ref.~\onlinecite{DagottoBook}.

To discuss the orbital pattern explicitly, let us focus on the
optimized configuration with model parameters $\lambda=1.5$,
$J^{\rm{AF}}=0.065$, and $\alpha=1.0$. For other $\alpha$ values the
patterns look very similar. The orbital patterns in layers 1 and 2,
where the bulk-like A-AFM phase is stabilized, display a clear
staggered $3x^2-r^2/3y^2-r^2$ order. Reciprocally, layers 7 and 8 do
not show any orbital order due to the vanishing value of the
electronic density. However, the orbital patterns close to the
interface in the range of layers from 3 to 6 are complicated and
they are presented in Fig.~\ref{F.OrbCE}. Here, we observe a
transition from the staggered $3x^2-r^2/3y^2-r^2$ order to the
orbital order of the CE state, and then to a (very weak) $3z^2-r^2$
order in the G-AFM state, with increasing layer index. In layer 3,
the orbital pattern is very close to that expected of a $n=1$ state,
but note that the orbital population along one of the orientations
of the diagonals is not identical for each diagonal. This is caused
by the influence of the CE state of the layer 4. In addition, in the
CE state of layers 4 and 5, the ``bridge" sites, which have a higher
electronic density than the rest, have staggered $3x^2-r^2/3y^2-r^2$
order, but the ``corner'' sites show a uniform $x^2-y^2$ order,
which is larger than in the CE state stabilized in the bulk. In
general, we find that in layers where either FM or CE spin order
exist the $e_{\rm g}$ electrons prefer to form an in-plane orbital
order. This is because in these layers the two oxygens connected to
a Mn ion tend to shrink along the $z$ direction to minimize the
energy of electron-phonon interactions by decreasing $Q_3$. But in
the layers where there is a G-AFM state, the oxygens will expand
along the $z$ direction to partially compensate the shrinking effect
along this direction in other layers.~\cite{footnote} Hence, a
$3z^2-r^2$ order may appear. In summary, the features of the orbital
arrangements are dominated by what we expect to find in $n=1$ A-AFM,
$n=0.5$ CE, and $n=0.0$ G-AFM states. However, subtle deviations can
be observed: non equivalent diagonals in layer 3, corner population
in the zigzags of layers 4 and 5, and weak orbital occupation along
the $z$-axis in layer 6.

\section{Results for wide bandwidth manganites}
In the previous section, the physical properties of the
$\rm{LaMnO_3/CaMnO_3}$ and $\rm{PrMnO_3/CaMnO_3}$ heterostructures were discussed.
States with CE characteristics were found to be stabilized near the interface. But for the cases of the
$\rm{LaMnO_3/SrMnO_3}$ and $\rm{PrMnO_3/SrMnO_3}$ heterostructures,
we expect to obtain different interfacial states
because the corresponding bulk materials, such as $\rm{(La,Sr)MnO_3}$
and $\rm{(Pr,Sr)MnO_3}$, have a wider bandwidth and the CE state is not
stabilized  in these compounds at half doping.

\subsection{The bulk phase diagram}
\begin{figure}
\begin{center}
\includegraphics[
clip,width=90mm,angle=0]{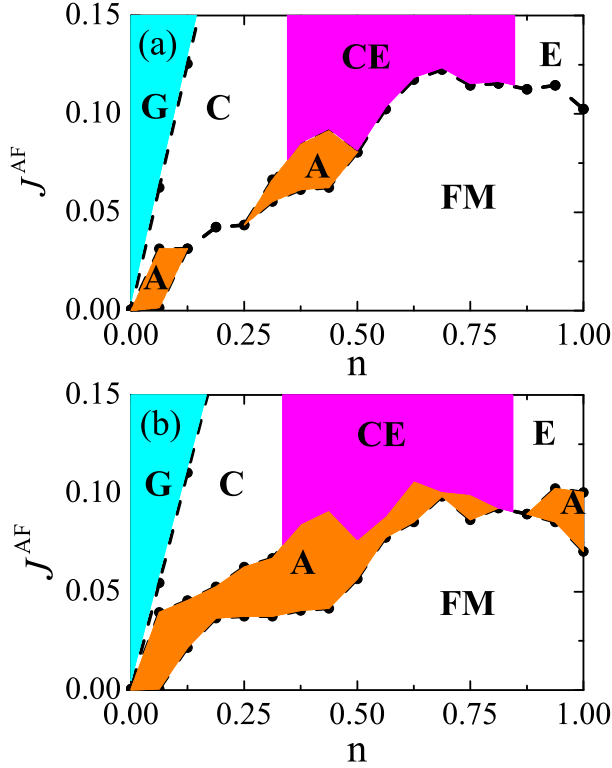} \vskip -0.5cm \caption{Phase
diagram of the two-orbital model on the $4\times4\times4$ lattice
with $\lambda=1.2$. Shown are the cases: (a) $J^{\rm{AF}}_\parallel
= J^{\rm{AF}}_\perp = J^{\rm{AF}}$, and (b) $J^{\rm{AF}}_\parallel =
2J^{\rm{AF}}_\perp/3 = J^{\rm{AF}}$. The results were obtained
comparing the energies of the states G-AFM, C-AFM, CE-AFM, A-AFM,
E-AFM, and FM.} \label{F.PhDAG}
\end{center}
\end{figure}

As in the previous section,
let us first study the phase diagram of the wide bandwidth
manganite in bulk form, focusing here on the case $\lambda=1.2$. This
coupling is smaller than the one used for narrow bandwidth manganites. However, for
$J^{\rm{AF}}_\parallel=J^{\rm{AF}}_\perp=J^{\rm{AF}}$, as shown in
Fig.~\ref{F.PhDAG}(a), the A-AFM phase at $\lambda=1.2$ cannot be stabilized near
$n\approx1$. This is because the A-AFM phase only appears in a very
narrow regime of the phase diagram in the two-orbital model if a
cubic lattice symmetry is considered.~\cite{Hotta03} This is compatible with the fact that the
real compounds exhibiting the A-AFM phase, such as $\rm{LaMnO_3}$,
have an orthorhombic, instead of cubic, lattice symmetry, with the
lattice constant along the $c$ axis smaller than those along the $a$ and
$b$ axes.~\cite{Rodriguez98} To better incorporate the lattice distortion found in
the bulk parent compound, an inter-layer
superexchange coupling larger than the intra-layer one should be used, i.e.
$J^{\rm{AF}}_\perp > J^{\rm{AF}}_\parallel$.~\cite{DongLMOSMO08,Hotta99} The
ratio $J^{\rm{AF}}_\perp/J^{\rm{AF}}_\parallel$ is
estimated~\cite{DongLMOSMO08,Hotta99,ZhouGoodenough08} to be $1.2\sim1.5$. Using these numbers,
the phase diagram for $J^{\rm{AF}}_\parallel =
2J^{\rm{AF}}_\perp/3 = J^{\rm{AF}}$ at $\lambda=1.2$ is shown in
Fig.~\ref{F.PhDAG}(b). The A-AFM phase is now stabilized in a
wider regime of the phase diagram, including $n\approx1$. Thus, tuning
$J^{\rm{AF}}$ to $0.07$,
the system experiences transitions involving the G-C-A-FM-A phases
with increasing electronic density $n$, and this resembles properly the
experimentally observed phases of
wide-bandwidth manganites $\rm{(R,A)MO}$.~\cite{Kajimotoetal02}

As for the $\rm{RMO/AMO}$ heterostructure, the state on each side,
far from the interface, must converge to its bulk phase. To
obtain a stable A-AFM phase in the bulk $\rm{RMO}$, we use the couplings
$\lambda=1.2$ and $J^{\rm{AF}}_\parallel = 2J^{\rm{AF}}_\perp/3 =
J^{\rm{AF}} = 0.07$ on the $\rm{RMO}$ side of the heterostructure.
On the $\rm{AMO}$ side, since the bulk $\rm{AMO}$ still has the
cubic lattice symmetry,~\cite{Bhattacharya07} we adopt $\lambda=1.2$
and $J^{\rm{AF}}_\parallel = J^{\rm{AF}}_\perp = J^{\rm{AF}} =
0.07$. At the interface, the lattice constants along the $a$ and $b$
axes will compress but those along the $c$ axis elongate to recover
the cubic symmetry due to the strain effect. Thus, we adopt the
inter-layer superexchange coupling $J^{\rm{AF}}_\perp = J^{\rm{AF}}
= 0.07$ at the interface.

\subsection{The state at the interface of the heterostructure}
\begin{figure}
\begin{center}
\includegraphics[
clip,width=90mm,angle=0]{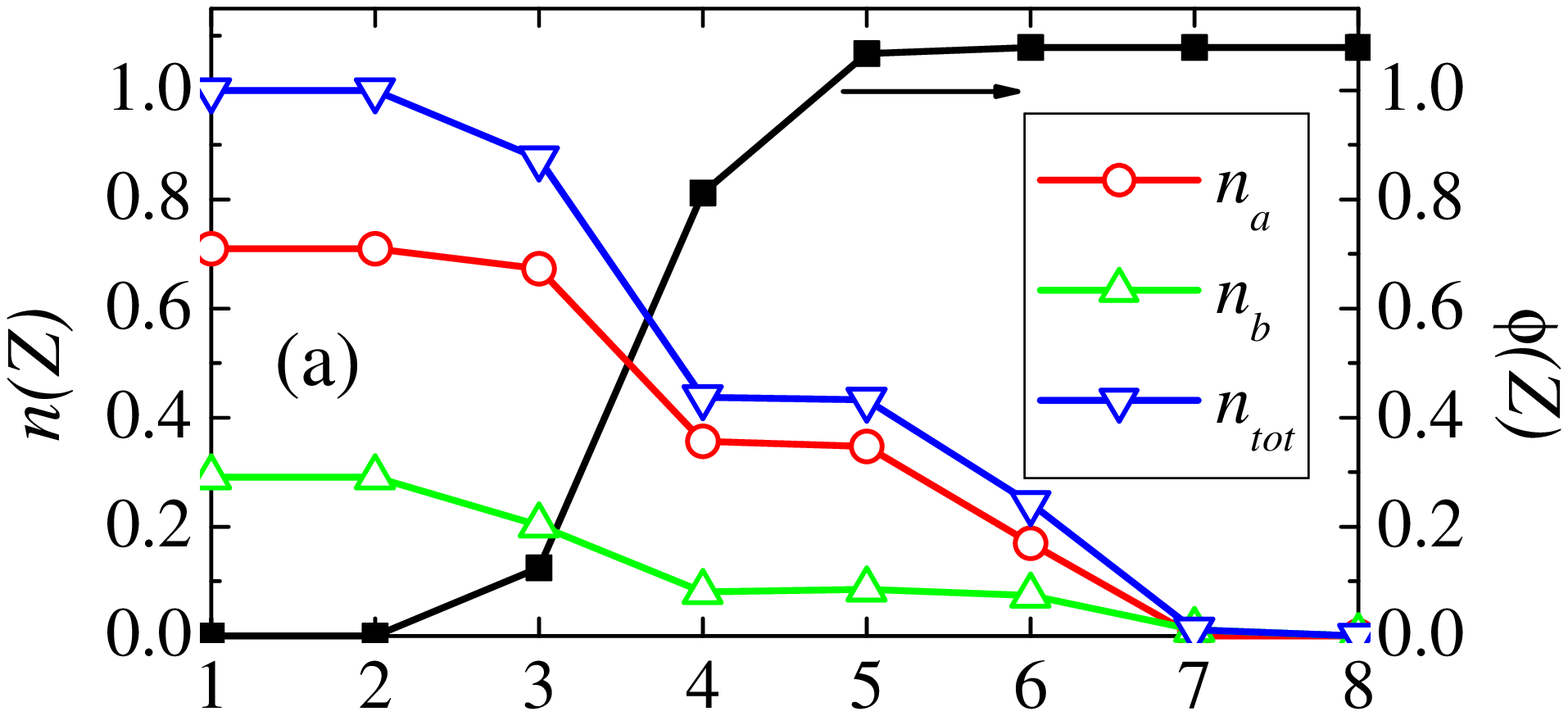} \vskip -0.95cm
\includegraphics[
clip,width=90mm,angle=0]{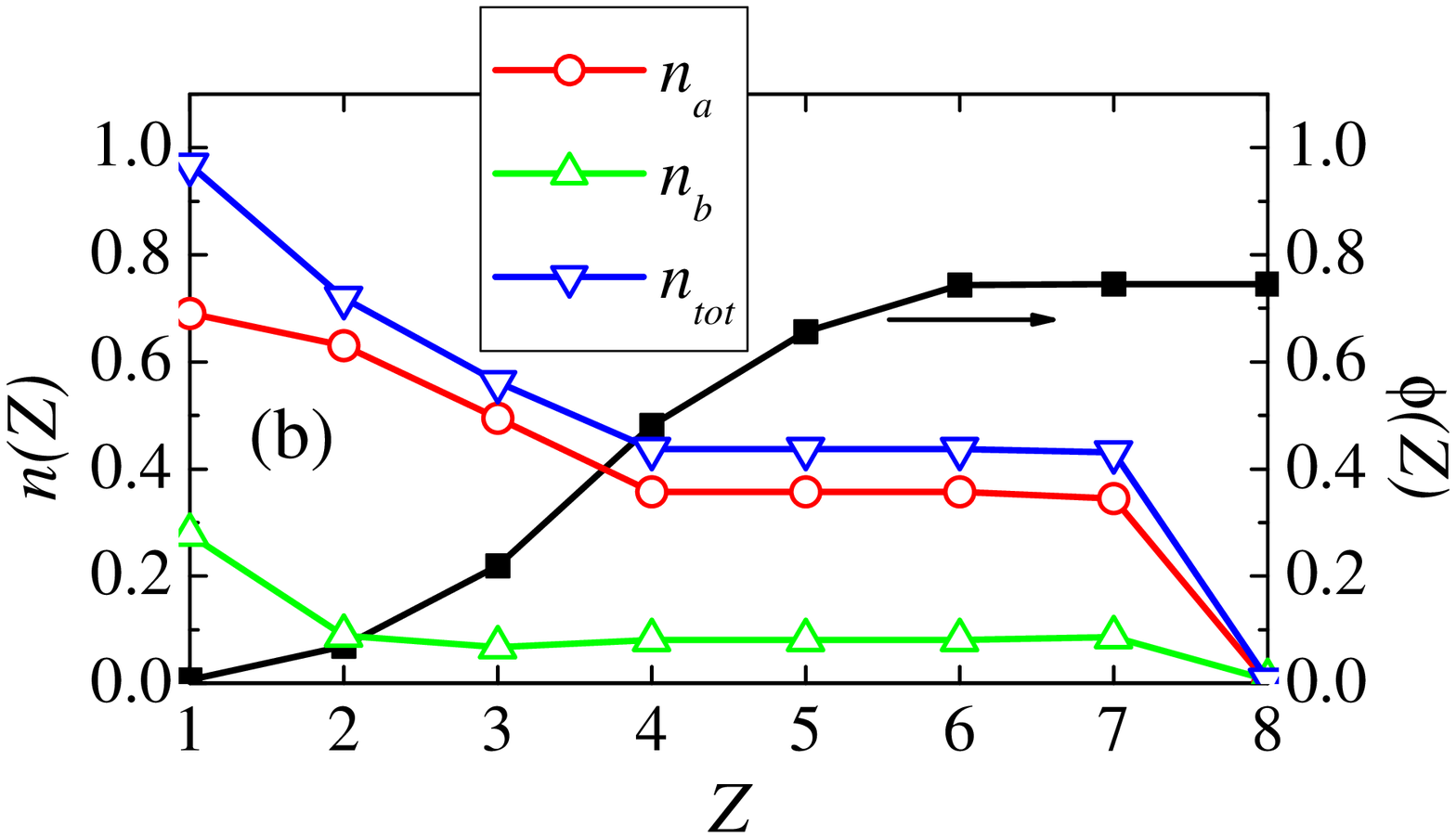} \vskip -0.5cm
\caption{Layer-averaged electron density $ n(Z)$ and electrostatic
potential $\phi(Z)$, obtained using $\lambda=1.2$ and $J^{\rm
AF}=0.07$. $n_a$ and $n_b$ refer to the electron densities of the
$a$ and $b$ orbitals, and $n_{\rm tot}=n_a+n_b$. (a) are results for
$\alpha=1.0$; (b) are results for $\alpha=0.2$.} \label{F.Lcdg1.22}
\end{center}
\end{figure}

\begin{figure}
\begin{center}
\includegraphics[
clip,width=80mm,angle=0]{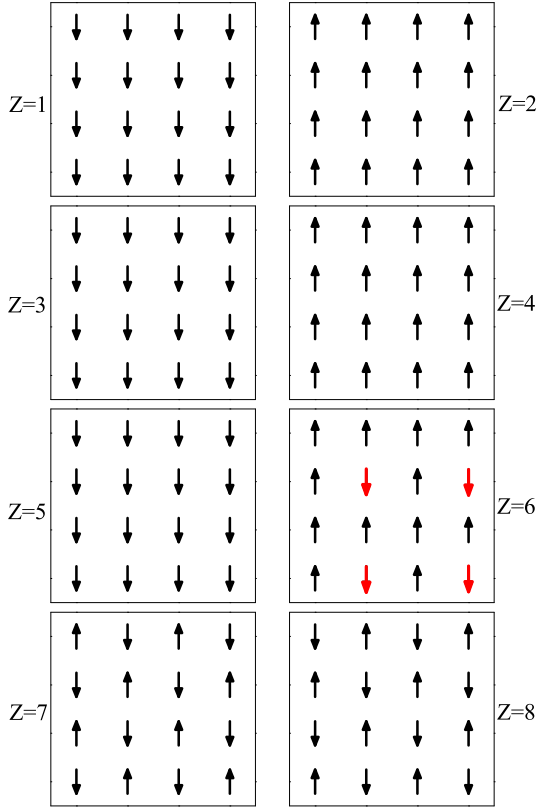} \caption{The optimized
real-space spin configuration at each layer of the heterostructure
with the model parameters used in Fig.~\ref{F.Lcdg1.22}(a).}
\label{F.Spng1.22a1.0}
\end{center}
\end{figure}

\begin{figure}
\begin{center}
\includegraphics[
clip,width=80mm,angle=0]{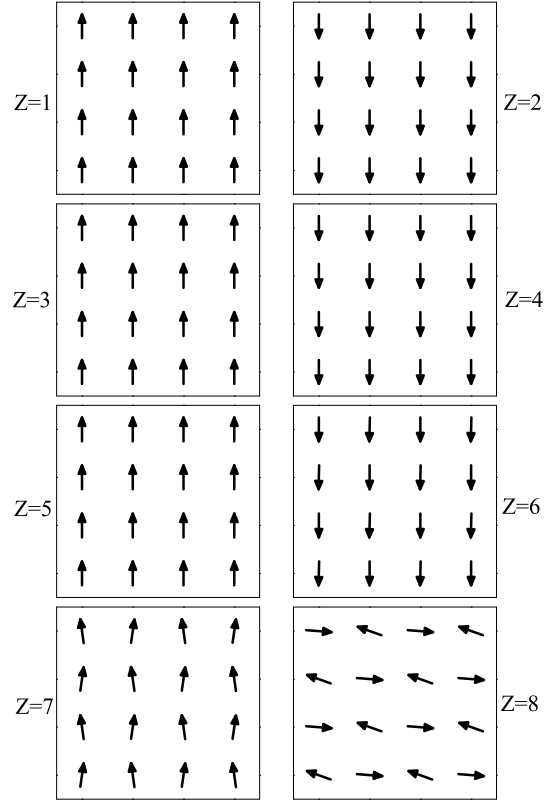} \caption{The optimized
real-space spin configuration at each layer of the heterostructure
with the model parameters used in Fig.~\ref{F.Lcdg1.22}(b).}
\label{F.Spng1.22a0.2}
\end{center}
\end{figure}

The averaged electrostatic potential and electronic densities in each
layer at $\alpha=1.0$ and $\alpha=0.2$ are presented in
Fig.~\ref{F.Lcdg1.22}. These charge distributions are actually
similar to those shown in Fig.~\ref{F.LcdCE}. Interestingly,
there is also a plateau at $n\approx0.5$ indicating the existence of a
fairly stable half-doped
state near the interface. Although there is no
CE phase in the phase diagram of wide bandwidth manganites, we find
instead that the A-AFM phase can be stabilized at $n\approx0.5$ in the bulk
limit. Thus, the plateau at $n\approx0.5$ in Fig.~\ref{F.Lcdg1.22}
suggests the state at the interface to be the A-AFM state. This assumption
is fully supported by the optimized real-space spin configuration results
presented in Figs.~\ref{F.Spng1.22a1.0} and \ref{F.Spng1.22a0.2}: for both $\alpha=1.0$ and $\alpha=0.2$,
the A-AFM phase spin arrangement is found in layers close to the interface.

Figures \ref{F.Spng1.22a1.0} and \ref{F.Spng1.22a0.2} indicate
that the transition from the A-AFM state at the interface to the G-AFM
state at one end of the heterostructure depends on the value of the
screening parameter $\alpha$. At a large $\alpha$ value, an interesting
result is found. In this case, there is one
layer (layer 6) with an intermediate exotic state, that has no analog in the bulk experimental
phase diagrams, to our
knowledge. This state consists of
alternate FM and AFM stripes, indicating a local mixed phase
tendency. Note that we have carried out
numerical studies on square clusters, $\lambda=1.2$, and $J^{\rm AF}=0.07$,
simulating bulk two dimensional systems, and in this case we do find
a similar mixed AF-FM state at electronic quarter-filling densities. Thus,
it is conceivable that this state may exist in bulk
single-layer manganites at large hole doping as well.
At small $\alpha$ values, on the other hand, the transition
is via introducing spin canting in several layers. We note that in
real heterostructures, the mixed-phase tendencies and spin canting may
coexist.

\begin{figure}
\begin{center}
\vskip -0.8cm
\includegraphics[
clip,width=80mm,angle=0]{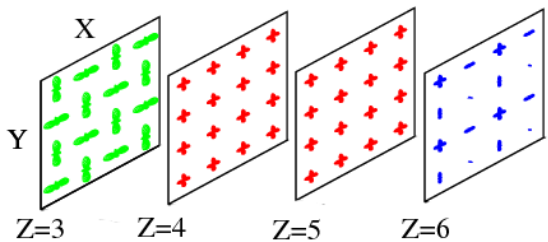} \vskip -0.8cm \caption{The
orbital pattern found in layers 3 to 6 of the heterostructure. A
uniform $x^2-y^2$ orbital order is observed at the interface. The
orbital order in layer 6 is exotic, i.e. not found in the bulk,
similarly as the spin arrangement of the same layer.}
\label{F.Orbg1.22}
\end{center}
\end{figure}

Let us consider now the orbital occupation near
the interface. In Fig.~\ref{F.Orbg1.22} the orbital pattern near the
interface for $\lambda=1.2$, $J^{\rm{AF}}=0.07$, and $\alpha=1.0$ is
presented. It is interesting to observe that the orbital pattern
shows a clear transition from a staggered $3x^2-r^2/3y^2-r^2$ order at one
end of the heterostructure to a uniform $x^2-y^2$ order at the
interface, layers 4 and 5, although the spin order is unchanged and fixed into an
A-AFM state. This is understandable, since we know in the bulk the A-AFM
exists both close to $n=1$ and close to $n=0.5$.
Comparing with Fig.~\ref{F.Lcdg1.22}(a), we observe that the orbital
order is tightly connected to the average electronic density of the
layer: the $3x^2-r^2/3y^2-r^2$ order appears at $n\approx1$, but the
$x^2-y^2$ order is present at $n\approx0.5$. Such a uniform $x^2-y^2$
orbital order is also observed in the A-AFM phase at $n\approx0.5$
in the bulk material.~\cite{DagottoBook,OrbOrderA} Since the bulk
A-AFM phase at $n\approx0.5$ is metallic, the A-AFM state at the
interface of the heterostructure can be anticipated to be a two-dimensional metal.
However, note that such a 2D metallic state could be unstable due to Anderson localization
introduced by the roughness and defects at the interface. Hence, an
insulating behavior is more likely to be observed in real materials.

In the exotic layer 6 exhibiting the mixed AFM-FM tendency, the orbital
pattern is complicated. According to Fig.~\ref{F.Spng1.22a1.0},
there are majority spins (black arrows in layer $6$) and minority spins
(red arrows). The minority spin sites correspond to $3z^2-r^2$ order. The
sites adjacent to the minority spin sites display alternating
$3x^2-r^2/3y^2-r^2$ order, but the sites diagonal to the minority spin
sites show a $x^2-y^2$ order. Once again, we remark that such an exotic orbital
order has not been observed experimentally in bulk materials to our knowledge,
although it may be part of theoretical phase diagrams of models for
two dimensional manganites
in the bulk at large hole densities, and it is also
conceivable that real single-layer
manganites in the bulk may present also a similar phase.

\section{Discussion and Conclusions}

In this paper, the magnetic and electronic
properties of the states near the interface of $\rm{RMO/AMO}$
heterostructures were investigated using numerical optimization techniques
on small clusters at zero temperature. The states stabilized at the interface
are found to
be similar to the states in the bulk compound at electronic density
$n\approx0.5$. This is easy to understand. Let us consider the
superlattice $\rm{(RMO)}_m\rm{/(AMO)}_n$. When ${\rm m,n}\gtrsim a/\alpha$
where $a$ is the lattice constant along the $z$ direction, there must
exist layers exhibiting properties of the bulk parent compounds
$\rm{RMO}$ and $\rm{AMO}$. In this case, the electronic density in the
regime near the interface is not much sensitive to $\rm m$ or $\rm n$, but is
always close to $0.5$ if $\alpha$ is about the same on both sides of
the heterostructure. Then, the regime close to the interface (within
$a/\alpha$ in the $z$ direction) can accommodate the state stabilized at
$n\approx0.5$ in the bulk. For the case of a wide bandwidth
manganite, a uniformly $x^2-y^2$ ordered A-AFM phase is
stabilized at the interface. Hence we propose that this state should
exist in the long-period $\rm{(LMO)_{\rm m}/(SMO)_{\rm n}}$ superlattice where
$m$ and $n$ are large enough such that there are bulk-like A-AFM and
G-AFM insulating regimes far from the interface. It is quite
reassuring that recent experiments on the $\rm{(LMO)_n/(SMO)_{2n}}$
superlattice at ${\rm n}=3$ provided strong evidence for the existence of this
orbital ordered A-AFM phase.~\cite{Bhattacharya07} However, for
short-period superlattices, the electronic density near the interface
may deviate from $0.5$, and also note that a well-defined
bulk-like regime may not exist in these structures. Hence the state
stabilized at the interface may be different from the one
in the long-period superlattice. For instance, in the
$\rm{(LaMnO_3)}_{2n}\rm{/(SrMnO_3)}_n$ superlattice with ${\rm n}<4$, the
electronic density near the interface is always higher than
$0.5$,~\cite{DongLMOSMO08} falling into the FM regime in the phase
diagram of bulk $\rm{LSMO}$. Hence, it is natural to observe a metallic
FM state stabilized at the
interface.~\cite{May07,DongLMOSMO08,Bhattacharya08,Smadici07,Adamoetal08}
Increasing the number of Sr layers, bulk-like insulating regimes
appear and the superlattice is driven through the MIT to be an
insulator. The electronic density at the interface is also
reduced by increasing the number of Sr layers. As shown in
Ref.~\onlinecite{DongLMOSMO08}, the state at the interface is still FM with a
$3z^2-r^2$ orbital order. However, further increasing the number
of Sr layers, the electronic density at the interface will approach
$0.5$. Then, the state at the interface becomes an A-AFM with a
uniform $x^2-y^2$ orbital order as discussed above, and the properties
of the heterostructure are dominated by bulk-like
regimes.~\cite{May07}

When studying the wide bandwidth manganites, we have used
$J^{\rm{AF}}_\perp > J^{\rm{AF}}_\parallel$. For consistency,
the same superexchange coupling ratio should be also used for the
calculation of narrow-to-intermediate bandwidth manganites. But the
A-AFM phase at $n=1$ is already stabilized in a cubic lattice for a large electron-phonon coupling.
Thus, changing the superexchange coupling ratio does not modify the phase diagram
crucially. Calculations have shown that setting
$J^{\rm{AF}}_\perp > J^{\rm{AF}}_\parallel$ also does not change the
results for the heterostructures. Hence, we have only presented here the results with
$J^{\rm{AF}}_\perp = J^{\rm{AF}}_\parallel$ in Sec.~\ref{S.LMM}. In this
case, interfaces dominated by CE-AFM characteristics have been observed.

\begin{figure}
\begin{center}
\vskip -0.5cm
\includegraphics[
clip,width=75mm,angle=0]{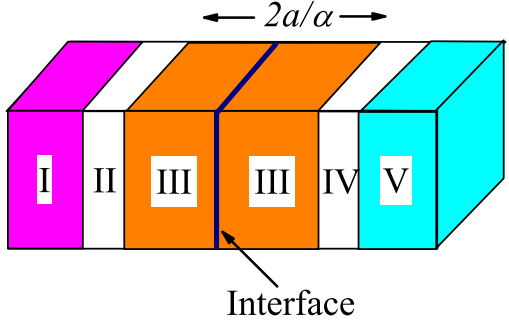} \caption{ Schematic
representation of the results found in our investigations. I and V
correspond to regions with properties similar to those of the bulk
of the two materials involved in the heterostructures. Region III is
very close to the interface. It is in this regime that the
electronic density is approximately 0.5 if the bulk components have
densities 1 and 0, as in our study. Depending on bandwidths here
either a CE or an A-AFM state are found. Finally, in regions II and
IV, the material must interpolate in properties between I and III or
III and V. These interpolations lead to states that do not appear to
have bulk analogs. } \label{F.Heterostruct2}
\end{center}
\end{figure}

It is very important to remark that several of our results have
unveiled phases that are not observed in the bulk phase diagrams.
They correspond to interesting modifications of the well-established
bulk phases. For instance, for narrow bandwidth manganite
heterostructures, the existence of exotic spin arrangements, such as
``canted CE'' and others, have been reported in our investigations.
For wide bandwidth manganites, unexpected mixtures of FM and AF
features were also identified. While our observations obtained on
small clusters need to be confirmed by other many-body techniques,
the present computational studies revealed the possibility of
finding new phases at interfaces, that do not exist in the bulk.
This is an exciting result that deserves further investigations.
Figure~\ref{F.Heterostruct2} summarizes schematically our results.

Regarding size effects in our simulations, certainly it is possible that the actual spin, charge,
and orbital patterns
of the novel states may be more complicated than found in our present study that was
limited to small systems. However, note that in the heterostructures discussed here the
$n=1$ A-AFM on one side, the $n=0.5$ CE or A-AFM at the interface, and the $n=0$ G-AFM on the other
side are very robust and likely will be present in real heterostructures. With
these states (with different layer electronic density)
anchored somewhere in the structure, then it is very reasonable
to expect magnetic
states $interpolating$ between them, that will have novel properties.
In the bulk when
the electronic concentration is the same in every layer of course it never
happens that a magnetic layer must interpolate between others with different densities. Thus, we
are confident that novel magnetic states, of a form likely even more complex than unveiled here, would
be present in manganite superstructures if simulations using larger clusters were possible.
Note that for bulk
phase-separated manganites,~\cite{DagottoReview01} involving a competition between states with different electronic densities,
then the novel states discussed here could also
appear at the interfaces between puddles of the competing phases as well.

In conclusion, we find that the properties of the $\rm{RMO/AMO}$
heterostructure are closely, but not entirely, associated with the
phase diagram of the bulk compound $\rm{(R,A)MO}$. We summarize our
main results in Fig.~\ref{F.Heterostruct2}. As one sees, although a
uniform state does not exist in the heterostructure due to the
redistribution of the charges, as a first approximation the state
near the interface can be ``read" from the phase diagram of the bulk
compound $\rm{(R,A)MO}$ at electronic density $n\approx0.5$, i.e., a
CE state if $\rm{(R,A)MO}$ is a narrow-to-intermediate bandwidth
manganite, but an A-AFM state with uniform $x^2-y^2$ orbital order
if $\rm{(R,A)MO}$ is a wide bandwidth manganite. However, the
intermediate states in between the interface and the bulk-like
regimes are sensitive to both the model parameters and the length of
the heterostructure. At least in our studies, it can be either a
spin canted CE state or a state showing local FM-AF mixed
tendencies. We believe our results can be used to describe the
ground state properties of the long-period $\rm{(RMO)}_{\rm
m}\rm{/(AMO)}_{\rm n}$ superlattices as well.

\section{Acknowledgment}

We thank L. Brey, M. Daghofer, C. Lin, S. May, D. Khomskii, S.
Okamoto, J. Salafranca, and Y. Tokura for useful discussions. This
work was supported by the NSF under Grant No. DMR-0706020 and the
Division of Materials Science and Engineering, U.S. DOE, under
contract with UT-Battelle, LLC. S.Y. is also supported in part by
CREST (JST).


\begin{thebibliography}{99}
\bibitem{Izumi99} M. Izumi, Y. Murakami, Y. Konishi, T. Manako, M.
Kawasaki, and Y. Tokura, Phys. Rev. B {\bf 60}, 1211 (1999).
\bibitem{Chakhalian08} J. Chakhalian, J. W. Freeland, H.-U.
Habermeier, G. Cristiani, G. Khaliullin, M. van Veenendaal, and B.
Keimer, Science {\bf 318}, 1114 (2008).

\bibitem{Okamoto04} S. Okamoto and A. J. Millis, Phys. Rev. B {\bf
70}, 241104(R) (2004); S. Okamoto and A. J. Millis, Nature {\bf
428}, 630 (2004).

\bibitem{Chaloupka08} J. Chaloupka and G. Khaliullin, Phys. Rev.
Lett. {\bf 100}, 016404 (2008).


\bibitem{Yunoki07} S. Yunoki, A. Moreo, E. Dagotto, S. Okamoto,
S. S. Kancharla, and A. Fujimori, Phys. Rev. B {\bf 76}, 064532
(2007).
\bibitem{Lin08} C. Lin and A. J. Millis, Phys. Rev. B {\bf 78}, 184405 (2008).
\bibitem{Adamoetal08} C. Adamo, X. Ke, P. Schiffer, A. Soukiassian,
M. Warusawithana, L. Maritato, and D. G. Schlom, Appl. Phys. Lett.
{\bf 92}, 112508 (2008).
\bibitem{Yamadaetal06} H. Yamada, M. Kawasaki, T. Lottermoser, T.
Arima, and Y. Tokura, Appl. Phys. Lett. {\bf 89}, 052506 (2006).
\bibitem{Nanda08} B. R. K. Nanda and S. Satpathy, Phys. Rev. B {\bf 78}, 054427 (2008).
\bibitem{May07} S. J. May, A. B. Shah, S. G. E. te Velthuis,
M. R. Fitzsimmons, J. M. Zuo, X. Zhai, J. N. Eckstein, S. D. Bader,
and A. Bhattacharya, Phys. Rev. B {\bf 77}, 174409 (2008).
\bibitem{Smadici07} \c{S}. Smadici, P. Abbamonte, A. Bhattacharya,
X. Zhai, B. Jiang, A. Rusydi, J. N. Eckstein, S. D. Bader, and J.-M.
Zuo, Phys. Rev. Lett. {\bf 99}, 196404 (2007).
\bibitem{Koida02} T. Koida, M. Lippmaa, T. Fukumura, K. Itaka, Y.
Matsumoto, M. Kawasaki, and H. Koinuma, Phys. Rev. B {\bf 66},
144418 (2002).
\bibitem{Bhattacharya07} A. Bhattacharya, X. Zhai, M. Warusawithana,
J. N. Eckstein, and S. D. Bader, Appl. Phys. Lett. {\bf 90}, 222503
(2007).
\bibitem{DongLMOSMO08} S. Dong, R. Yu, S. Yunoki, G. Alvarez, J.-M. Liu,
and E. Dagotto, Phys. Rev. B {\bf 78}, 201102(R) (2008).
\bibitem{Bhattacharya08} A. Bhattacharya, S. J. May, S. G. E. te Velthuis, M. Warusawithana, X. Zhai, B. Jiang, J. M. Zuo, M. R. Fitzsimmons, S. D. Bader, and J. N. Eckstein, Phys. Rev. Lett. {\bf 100}, 257203 (2008).
\bibitem{Yunoki08} S. Yunoki, E. Dagotto, S. Costamagna, and J. A.
Riera, Phys. Rev. B {\bf 78}, 024405 (2008); I. Gonz\'{a}lez, S.
Okamoto, S. Yunoki, A. Moreo, and E. Dagotto, J. Phys.: Condens.
Matter {\bf 20}, 264002 (2008).

\bibitem{Brey07} L. Brey, Phys. Rev. B {\bf 75}, 104423 (2007).

\bibitem{Brey08} M. J. Calder\'{o}n, J. Salafranca, and L. Brey, Phys. Rev. B {\bf
78}, 024415 (2008); J. Salafranca,  M. J. Calder\'{o}n, and L. Brey,
Phys. Rev. B {\bf 77}, 014441 (2008).

\bibitem{Lin06} C. Lin, S. Okamoto, and A. J. Millis, Phys. Rev. B
{\bf 73}, 041104 (2006).

\bibitem{DagottoReview01} E. Dagotto, T. Hotta, and A. Moreo, Phys.
Rep. {\bf 344}, 1 (2001).

\bibitem{Hotta99} T. Hotta, S. Yunoki, M. Mayr, and E. Dagotto, Phys.
Rev. B {\bf 60}, 15009(R) (1999).

\bibitem{NumRec} W. H. Press, S. A. Teukolsky, W. T. Vetterling,
and B. P. Flannery, \emph{Numerical Recipes}, Vol. 1 (\emph{2nd Ed.},
Cambridge University Press, 1992).

\bibitem{Hotta03} T. Hotta, M. Moraghebi, A. Feiguin, A. Moreo,
S. Yunoki, and E. Dagotto, Phys. Rev. Lett. {\bf 90}, 247203 (2003).

\bibitem{spiral} S. Dong, R. Yu, S. Yunoki, J.-M. Liu, and E. Dagotto.  Phys. Rev. B {\bf 78}, 155121 (2008).

\bibitem{bulkAphase} Y. Murakami, J. P. Hill, D. Gibbs, M. Blume,
I. Koyama, M. Tanaka, H. Kawata, T. Arima, Y. Tokura, K. Hirota,
and Y. Endoh, Phys. Rev. Lett. {\bf
81}, 582 (1998).


\bibitem{Kajimotoetal02} R. Kajimoto, H. Yoshizawa, Y. Tomioka,
and Y. Tokura, Phys. Rev. B {\bf 66}, 180402(R) (2002).


\bibitem{note_spin_order} Although the spin order is not sensitive
to the value of $\alpha$, the exact spin pattern does. This can be
best seen from Fig.~\ref{F.Spng1.22a1.0} and
Fig.~\ref{F.Spng1.22a0.2}, where the relative angles between spins
in layer 1 and layer 8 is different at different $\alpha$.

\bibitem{commentbrey} We remark that L. Brey in Ref.~\onlinecite{Brey07}
also previously reported a charge-ordered CE state at the interface between a FM metal and a large-gap
insulator, such as SrTiO$_3$, due to a
reduction of the electronic density at that interface.


\bibitem{cox} D. E. Cox, P. G. Radaelli, M. Marezio, and S-W. Cheong, Phys. Rev. B {\bf 57}, 3305 (1998). See also
C. Yaicle, C. Martin, Z. Jirak, F. Fauth, G. Andri, E. Suard, A. Maignan,
V. Hardy, R. Retoux, M. Hervieu, S. Hibert, B. Raveau, Ch. Simon, D. Saurel, A. Brulet, and F. Bouree,
Phys. Rev. B {\bf 68}, 224412 (2003).

\bibitem{yoshizawa} H. Yoshizawa, H. Kawano, Y. Tomioka, and Y. Tokura, Phys. Rev. B {\bf 52}, R13145 (1995).

\bibitem{radaelli} P. G. Radaelli, R. M. Ibberson, D. N. Argyriou, H. Casalta, K. H. Andersen, S.-W. Cheong, and J. F. Mitchell,
Phys. Rev. B {\bf 63}, 172419 (2001), and references therein. See also page 292 of Ref.~\onlinecite{DagottoBook}.

\bibitem{khomskii} In D. Efremov, J. van den Brink, and D. Khomskii, Nature Mat. {\bf 3}, 853 (2004), a modified CE state was discussed in the context of novel ferroelectric states. In this reference, the orientation
of the spins within the zigzag chains is modified with respect to the pattern in the
perfect CE state via the introduction of
dimers. The state of Efremov {\it et al.} does not appear to be
the same as in the ``canted CE'' state discussed in the present paper, but
further work is needed to fully address this issue.



\bibitem{DagottoBook} E. Dagotto, \emph{Nanoscale Phase Separation
and Colossal Magnetoresistance}, Springer Verlag, Berlin, 2002.

\bibitem{footnote} The compensation is not complete when considering the breathing mode
in the model. In heterostructures, the periodicity of the lattice distortions along
the $z$ direction is lost so that the total volume of the lattice
may not be kept constant due to the existence of the breathing mode.


\bibitem{Rodriguez98} J. Rodr\'{i}guez-Carvajal, M. Hennion, F. Moussa,
A. H. Moudden, L. Pinsard, and A. Revcolevschi, Phys. Rev. B {\bf
57}, R3189 (1998).


\bibitem{ZhouGoodenough08} J. Zhou and J. Goodenough, Phys. Rev. B
{\bf 77}, 132104 (2008).
\bibitem{OrbOrderA} R. Kajimoto, H. Yoshizawa, H. Kawano, H. Kuwahara, Y.
Tokura, K. Ohoyama, and M. Ohashi, Phys. Rev. B {\bf 60}, 9506
(1999); A. Urushibara, Y. Moritomo, T. Arima, A. Asamitsu, G. Kido,
and Y. Tokura, Phys. Rev. B {\bf 51}, 14103 (1995); Y. Tokura, A.
Urushibara, Y. Moritomo, T. Arima1, A. Asamitsu, G. Kido and N.
Furukawa, J. Phys. Soc. Jpn. {\bf 63}, 3931 (1994).



\end{thebibliography}
\end{document}